%Paper: gr-qc/9303006
%From: hayward@murasaki.scphys.kyoto-u.ac.jp
%Date: Tue, 2 Mar 93 16:27:53 +0100
%Date (revised): Mon, 29 Mar 93 16:51:54 +0200
%Date (revised): Wed, 25 May 94 15:08:20 JST

\font\lbf=cmbx10 scaled\magstep2

\def\bs{\bigskip}
\def\ms{\medskip}
\def\np{\vfill\eject}

\def\ni{\noindent}
\def\cl{\centerline}

\def\title#1{\cl{\lbf #1}}

\def\ref#1#2#3#4{#1\ {\it#2\ }{\bf#3\ }#4\par}
\def\refb#1#2#3{#1\ {\it#2\ }#3\par}
\def\CQG{Class.\ Qu.\ Grav.}

\def\PR{Phys.\ Rev.}

\def\con#1#2{\left\langle#1,#2\right\rangle}
\def\O#1{\left.#1\right\vert_S}
\def\H#1{\left.#1\right\vert_H}
\def\I{\int_S\mu}
\def\C{\overline{H}}
\def\D{{\cal D}}
\def\E{{\cal E}}
\def\L{{\cal L}}
\def\M{{\cal M}}
\def\R{{\cal R}}
\def\S{{\cal S}}
\def\U{{\cal U}}
\def\N{\nabla}
\def\a{\alpha}
\def\b{\beta}
\def\g{\gamma}
\def\d{\delta}
\def\e{\varepsilon}
\def\k{\kappa}

\def\o{\omega}
\def\s{\sigma}
\def\t{\theta}
\def\half{{\textstyle{1\over2}}}
\def\quart{{\textstyle{1\over4}}}
\def\tt{{\textstyle{1\over32}}}

\magnification=\magstep1

\title{General laws of black-hole dynamics}
\bs\cl{\bf Sean A. Hayward}
\ms\cl{Max-Planck-Institut f\"ur Astrophysik}
\cl{Karl-Schwarzschild-Stra\ss e 1}
\cl{85740 Garching bei M\"unchen}
\cl{Germany}
\ms\cl{Faculty of Mathematical Studies}
\cl{University of Southampton}
\cl{Southampton SO9 5NH}
\cl{United Kingdom}
\ms\cl{Current address:}
\cl{Department of Physics}
\cl{Kyoto University}
\cl{Kyoto 606-01}
\cl{Japan}
\bs\cl{Revised 6th April 1994}
\bs\ni
{\bf Abstract.}
A general definition of a black hole is given,
and general `laws of black-hole dynamics' derived.
The definition involves something similar to an apparent horizon,
a trapping horizon,
defined as a hypersurface foliated by marginal surfaces
of one of four non-degenerate types,
described as future or past, and outer or inner.
If the boundary of an inextendible trapped region is suitably regular,
then it is a (possibly degenerate) trapping horizon.
The future outer trapping horizon provides the definition of a black hole.
Outer marginal surfaces have spherical or planar topology.
Trapping horizons are null
only in the instantaneously stationary case,
and otherwise outer trapping horizons are spatial
and inner trapping horizons are Lorentzian.
Future outer trapping horizons have non-decreasing area form,
constant only in the null case---the `second law'.
A definition of the trapping gravity of an outer trapping horizon is given,
generalizing surface gravity.
The total trapping gravity of a compact outer marginal surface
has an upper bound, attained if and only if
the trapping gravity is constant---the `zeroth law'.
The variation of the area form along an outer trapping horizon
is determined by the trapping gravity and an energy flux---the `first law'.
\np\ni
{\bf I. Introduction}
\ms\ni
Perhaps the strongest evidence for a fundamental connection
between quantum physics and gravity
is the link between the laws of thermodynamics
and the analogous `laws of black-hole dynamics' [1].
Of the latter, the first and zeroth laws apply only to stationary black holes,
and the only law established in any generality is the second law,
Hawking's area theorem,
namely that the area of the event horizon cannot decrease [2--3].
Even this is not particularly general,
since the event horizon is
defined only in asymptotically flat space-times [2--4].
Indeed, black holes are conventionally defined
only in asymptotically flat space-times,
by the existence of an event horizon.
In practice, the universe is not thought to be asymptotically flat,
in which case event horizons do not exist.
Even if the universe were asymptotically flat,
mortal observers could not verify such a global property,
nor detect event horizons.
Generalization, both of the concept of black hole and the corresponding laws,
involves not the event horizon
but something similar to the apparent horizon [2--3],
introduced subsequently as the {\it trapping horizon},
which provides the operational definition of a black hole
in the sense that it can be detected by observers.
In this article,
general zeroth, first and second laws of black-hole dynamics
are established for trapping horizons in arbitrary space-times.
\ms
The context is the general theory of relativity.
The main concepts are introduced in Section III,
to which one may skip on a first reading.
The required geometry is described in Section II.
The relevant dynamical equations and energy inequalities
are given in Section IV.
The second law is established in Section V,
following results concerning the topology and signature of trapping horizons.
The zeroth and first laws are established in Section VI
in terms of a generalization of surface gravity, {\it trapping gravity}.
\bs\ni
{\bf II. Double-null foliations}
\ms\ni
The geometrical objects most relevant to the dynamics of trapping horizons
are spatial 2-surfaces embedded in space-time,
and foliations of such 2-surfaces.
The geometry of embedded 2-surfaces is less well known than
that of embedded 3-surfaces,
which is described by the `3+1' formalism.
The corresponding `2+2' formalism [5] takes a particularly simple form
in terms of the two unique null directions normal to a spatial 2-surface.
The resulting `double-null' or `dual-null' formalism [6],
describing two foliations of null 3-surfaces intersecting in spatial
2-surfaces,
is reviewed as follows,
with slightly different notation and additional explanatory comments.
\ms
Given a 4-manifold $\M$ with a metric $g$
of Lorentzian $({-}{+}{+}{+})$ signature,
consider a smooth embedding $\varphi:\S\times\U_+\times\U_-\to\M$,
where $\S$ is a 2-manifold
and $\U_\pm=[0,U_\pm)$ are half-open intervals.
Taking parameters $\xi_\pm\in\U_\pm$,
this generates two foliations of 3-surfaces $\Sigma_\pm$
given by constant $\xi_\pm$,
intersecting in a foliation of 2-surfaces.
A particular pair of foliations determines the parameters $\xi_\pm$
only up to diffeomorphisms $\xi_\pm\mapsto\zeta_\pm(\xi_\pm)$,
which relabel $\Sigma_\pm$.
Introduce the evolution vectors $u_\pm=\partial/\partial\xi_\pm$,
assumed future-pointing,
and the normal 1-forms $n_\pm=-\hbox{d}\xi_\pm$.
The minus sign ensures that the dual vectors $N_\pm=g^{-1}(n_\pm)$
are also future-pointing.
By construction:
$u_\pm$ commute, $[u_+,u_-]=0$;
$n_\pm$ are closed, $\hbox{d}n_\pm=0$;
$u_\pm$ are tangent to $\Sigma_\mp$, $n_\mp(u_\pm)=0$;
and $n_\pm$ are normalized with respect to $u_\pm$, $n_\pm(u_\pm)=-1$.
Assume now that $\Sigma_\pm$ are null,
$g(N_\pm,N_\pm)=0$.
Define the normalization $\hbox{e}^f=-g(N_+,N_-)$,
the induced 2-metric $h=g+2\hbox{e}^{-f}n_+\otimes n_-$,
and the shift 2-vectors $r_\pm=h^{-1}h(u_\pm)$.
Then the evolution vectors are related to the normals by
$u_\pm-r_\pm=\hbox{e}^{-f}N_\mp$,
and the metric may be written as
$$g=\pmatrix{
h(r_+,r_+)&h(r_+,r_-)-\hbox{e}^{-f}&h(r_+)\cr
h(r_+,r_-)-\hbox{e}^{-f}&h(r_-,r_-)&h(r_-)\cr
h(r_+)&h(r_-)&h\cr}\eqno(1)$$
in a basis $(u_+,u_-;\hbox{\bf e})$, where {\bf e} is a basis for $\S$.
This metric is clearly the most general consistent with $\Sigma_\pm$ being
null,
so that all such double-null foliations are included.
\ms
The Lie derivative along a vector $v$ is denoted by $\L_v$,
and the Lie derivative along the normal directions $u_\pm-r_\pm$
is shortened to $\L_\pm$.
The extrinsic curvature of the foliation may be encoded in
the expansions $\t_\pm$, shears $\s_\pm$, inaffinities $\nu_\pm$
and anholonomicity (or twist) $\o$,
defined by
$$\eqalignno
{&\t_\pm=\half h^{cd}\L_\pm h_{cd}&(2a)\cr
&\s^\pm_{ab}=h_a^ch_b^d\L_\pm h_{cd}
-\half h_{ab}h^{cd}\L_\pm h_{cd}&(2b)\cr
&\nu_\pm=\L_\pm f&(2c)\cr
&2\hbox{e}^{-f}\o_a=h_{ab}[u_--r_-,u_+-r_+]^b
=h_{ab}\L_-(u_+-r_+)^b=-h_{ab}\L_+(u_--r_-)^b.&(2d)\cr}$$
In the `2+2' Hamiltonian formulation [6],
$(h,f,r_\pm)$ are the `$q$s'
and $(\t_\pm,\s_\pm,\nu_\pm,\o)$ the independent `$p$s'.
Initial data for the vacuum Einstein equations may be given as
$r_+|_V$ on the patch $V=\varphi(\S\times\U_+\times\U_-)$,
$(\s_+,\nu_+)|_{S_+}$
on the null 3-surface $S_+=\varphi(\S\times\U_+\times\{0\})$,
$(\s_-,\nu_-,r_-)|_{S_-}$
on the null 3-surface $S_-=\varphi(\S\times\{0\}\times\U_-)$,
and $(h,f,\o,\t_+,\t_-)|_S$
on the spatial 2-surface $S=\varphi(\S\times\{0\}\times\{0\})$.
Of these,
$(f|_S,\nu_\pm|_{S_\pm})$
encode the freedom to choose different double-null foliations based on $S$,
while the shifts $(r_+|_V,r_-|_{S_-})$
encode the freedom to identify the points of the 2-surfaces in different ways
under $u_\pm$.
This leaves as free data $h|_S$, up to diffeomorphisms of $S$,
and $(\t_\pm|_S,\s_\pm|_{S_\pm},\o|_S)$.
In terms of classical geometrical quantities [7],
$\t_\pm$ and $\s_\pm$ are the trace and trace-free parts
of the second fundamental forms
$I\!\!I_\pm=\half\hbox{e}^f(\t_\pm h+\s_\pm)$,
and $\o$ is an invariant which replaces
the foliation-dependent normal fundamental forms
$\b_\pm=-\hbox{e}^f(\half\D f\pm\o)$,
as is explained in Appendix~B.
%The second and normal fundamental forms
%describe the extrinsic curvature of a single submanifold $S$,
%whereas the extrinsic curvature of a foliation of submanifolds
%contains additional information.
\ms
Concerning the intrinsic geometry of $S$, the 2-metric $h$ determines
a Ricci scalar $\R$, a covariant derivative $\D$
and an area 2-form $\mu$.
%(In coordinates, $\mu=\sqrt{\det h}\,\hbox{d}x^1\wedge\hbox{d}x^2$).
The area form defines a measure $\I$ on $S$.
\ms
Consider a particular double-null foliation,
i.e.\ a particular set of null 3-surfaces $\Sigma_\pm$.
Recall that the dynamical variables $(h,f,r_\pm)$
and $(\t_\pm,\s_\pm,\nu_\pm,\o)$ are determined only up to relabellings
$\xi_\pm\mapsto\zeta_\pm(\xi_\pm)$
and interchange $\xi_\pm\mapsto\xi_\mp$ of $\Sigma_\pm$.
Quantities invariant under relabellings and interchange
are true geometrical invariants of the double-null foliation.
More generally,
if a quantity $X$ transforms as $X\mapsto(\zeta_+')^a(\zeta_-')^bX$
under relabellings, then $X$ will be said to have {\it weight} $[+]^a[-]^b$.
The weights of all the quantities
which occur in the double-null form of the vacuum Einstein equations [6]
are as follows.
Invariants, i.e.\ those of weight 1,
are $h$ (and hence $\D$ and $\R$), $\o$ and $\D f$.
Those of weight $[\pm]^{-1}$ are
$r_\pm$, $\t_\pm$, $\s_\pm$, $\L_\pm h$, $\L_\pm\o$ and $\D\nu_\pm$.
Those of weight $[\pm]^{-2}$ are
$(\L_\pm+\nu_\pm)\t_\pm$.
Those of weight $[+]^{-1}[-]^{-1}$ are
$\hbox{e}^{-f}$,
$\L_\pm r_\mp$, $\L_\pm\t_\mp$, $\L_\pm\s_\mp$, $\L_\pm\nu_\mp$.
In particular, the following combinations are invariant:
$\hbox{e}^f\t_+\t_-$,
$\hbox{e}^f\s_+\otimes\s_-$,
$\hbox{e}^f\t_+\s_-$,
$\hbox{e}^f\t_-\s_+$,
$\hbox{e}^f\L_\pm\t_\mp$,
$\hbox{e}^f\L_\pm\s_\mp$,
$\hbox{e}^f\L_\pm\nu_\mp$.
This is important for the applications,
which involve a 1-parameter foliation of spatial 2-surfaces
generating a 3-surface $\Sigma$
which may have spatial, null or Lorentzian signature.
A neighbouring double-null foliation based on the 2-surfaces always exists,
and is unique if $\Sigma$ is not null.
Quantities such as $\hbox{e}^f\t_+\t_-$ and $\hbox{e}^f\L_\pm\t_\mp$
are then uniquely determined.
\bs\ni
{\bf III. Marginal surfaces and trapping horizons}
\ms\ni
A key concept concerning black holes
is the {\it trapped surface} in the sense of Penrose~[4]:
a compact spatial 2-surface $S$ on which $\O{\t_+\t_-}>0$,
where $\t_\pm$ are the expansions
in the future-pointing null directions normal to $S$.
The space-time is assumed time-orientable,
so that `future' and `past' can be assigned consistently.
One may then distinguish between a past trapped surface,
for which $\O{\t_\pm}>0$,
and a future trapped surface, for which $\O{\t_\pm}<0$.
These are associated with white and black holes respectively.
In the latter case, the idea is that
both the ingoing and the outgoing light rays are converging,
so that all signals from the surface are trapped inside a shrinking region.
Future trapped surfaces do not quite characterize black holes,
as can be seen from the Carter black-hole solutions
with positive cosmological constant [1],
in which future trapped surfaces exist not just in the black-hole region,
but also in the `cosmological' region neighbouring past infinity.
\ms
A {\it trapping boundary} is defined in Appendix A
as a boundary of an inextendible trapped region,
namely an inextendible region for which each point lies on some trapped
surface.
Less generally, one might assume a foliation of spatial 3-surfaces $\Sigma$,
and define a trapping boundary
as generated by the boundaries of trapped subsets of each $\Sigma$,
namely subsets of $\Sigma$
for which each point lies on some trapped surface lying in $\Sigma$.
This restricts to those trapping boundaries
for which the trapped region can be foliated in this way,
probably a physically reasonable assumption
for a suitable choice of foliation,
though certainly not for an arbitrary choice of foliation.
This is similar to Hawking's definition of apparent horizon [2--3],
except that Hawking imposed additional global assumptions,
namely that the space-time is `regular predictable',
a much more serious restriction.
In particular, the apparent horizon, like the event horizon,
is defined only in asymptotically flat space-times.
Proposition 9.2.9 of Hawking \& Ellis [3] asserts that
an apparent horizon is foliated by surfaces
on which one expansion vanishes,
sometimes called marginal surfaces.
This suggests replacing the concept of apparent horizon or trapping boundary
with a more operational definition directly in terms of marginal surfaces.
\ms
An immediate problem is that a marginal surface
does not precisely capture the idea
of the outer boundary of a black or white hole.
Firstly, there are `pathological' examples such as
plane-wave space-times or the de~Sitter space-time,
where each point has a marginal surface passing through it.
Secondly, in such examples as the Kerr or Reissner-Nordstr\"om black hole,
marginal surfaces exist not just on the outer horizons
but also on the inner horizons.
These are usually distinguished by assigning an `outward' direction
as determined by the global assumption of asymptotic flatness [2--3],
again a serious restriction.
A simpler resolution is suggested by noting that the inner and outer horizons
of the Kerr or Reissner-Nordstr\"om black hole can be distinguished by
whether the vanishing expansion increases or decreases across the horizon.
\ms\ni
{\it Definitions.}
A {\it marginal surface} is a spatial 2-surface $S$
on which one null expansion vanishes, fixed henceforth as $\O{\t_+}=0$.
A {\it trapping horizon} is the closure $\C$ of a 3-surface $H$
foliated by marginal surfaces on which
$\H{\t_-}\not=0$ and $\H{\L_-\t_+}\not=0$,
where the double-null foliation is adapted to the marginal surfaces.
The trapping horizon and marginal surfaces are said to be
{\it outer} if $\H{\L_-\t_+}<0$,
{\it inner} if $\H{\L_-\t_+}>0$,
{\it future} if $\H{\t_-}<0$ and
{\it past} if $\H{\t_-}>0$.
\ms\ni
The outer marginal surface should not be confused with
Hawking's `marginally outer trapped surface' [2--3],
for which the global assumption of asymptotic flatness is used
to assign an `outward' direction.
\ms
The most relevant case in the context of black holes
is the {\it future outer trapping horizon}.
In this case, the definition captures the idea that
the ingoing light rays should be converging, $\H{\t_-}<0$,
and the outgoing light rays should be instantaneously parallel on the horizon,
$\H{\t_+}=0$,
and diverging just outside the horizon and converging just inside,
$\H{\L_-\t_+}<0$.
This is the principal intuitive justification for the definition.
One could also consider degenerate cases such as
$\H{\t_+}=0$, $\H{\L_-\t_+}=0$, $\H{\L_-\L_-\t_+}<0$.
Apart from the exclusion of degenerate horizons,
the future outer trapping horizon provides a general {\it definition}
of a black hole.
The point is that black holes can be characterized
not by the mere existence of future trapped surfaces,
but by the onset of their formation,
as identified by future outer marginal surfaces.
\ms
According to Appendix A,
a suitably regular trapping boundary is composed of marginal surfaces $S$
on which $\O{\t_-}$ and $\O{\L_-\t_+}$ do not change sign.
If either sign varies,
the marginal surface does not lie in a trapping boundary,
and so can be excluded from consideration
in the context of black or white holes.
Conversely, a compact, future or past, outer or inner marginal surface
lies in the closure of a trapped region.
So it is only these four types of marginal surface
which are relevant to black or white holes.
In particular,
any spatial 2-surface sufficiently close to, and to the future of,
a compact future outer marginal surface, is a future trapped surface.
Thus the existence of such a marginal surface
can replace that of a trapped surface in, for instance,
the singularity theorems [3--4].
\ms
Inner trapping horizons also include the `cosmological horizons'
that occur in examples such as
Robertson-Walker cosmological space-times,
or in the Carter cosmological black-hole space-times [1].
This resolves the outstanding question of distinguishing
trapped surfaces of a `cosmological' type
from those which form in localized gravitational collapse to a black hole.
Regarding this question,
Tipler [8] suggested a closely related definition
of `non-cosmological trapped surfaces'
which are acausally connectable to a marginal surface in a certain way,
which turns out to ensure that
the marginal surface is of the (possibly degenerate) future outer type.
\ms
The differing signs of $\O{\t_+}$ and $\O{\t_-}$
ensure that $S$ is orientable, and determine a preferred orientation,
with a direction normal to $S$ being called {\it inward}
if it points towards the neighbouring trapped surfaces,
and {\it outward} if it points in the opposite sense.
Such an intrinsically preferred orientation
renders global assumptions unnecessary in distinguishing inside from outside.
Indeed, the definition of future or past, outer or inner trapping horizon
requires no global assumptions and depends only on the quasi-local geometry,
as is appropriate for realistic gravitational collapse.
In particular, the definition can be used in the cosmological context.
\bs\ni
{\bf IV. The focussing equations}
\ms\ni
The vacuum Einstein equations,
adapted to double-null foliations of 2-surfaces,
have been derived in a Hamiltonian form [6].
The only components of the Einstein equations required here are those
which concern the development of the expansions,
which take the dimensionless form\footnote\ddag
{For those more familiar with the spin-coefficient formalisms [9],
assuming a double-null foliation corresponds to
$0=\k=\rho-\bar\rho=\e+\bar\e=\tau-\bar\a-\b$,
plus their `primed' versions.
Apart from the lack of direct analogues of $\nu_\pm$,
the focussing equations correspond to
(4.12.32a), (4.12.32f) and their `primed' versions,
using (4.14.20) to eliminate $\Psi_2$.}
$$\eqalignno
{&\L_\pm\t_\pm+\nu_\pm\t_\pm+\half\t_\pm^2+\quart\s^\pm_{ab}\s_\pm^{ab}
=-8\pi\phi_\pm&(3a)\cr
&\L_\pm\t_\mp+\t_+\t_-+\hbox{e}^{-f}\left(
\half\R-(\half\D_af\pm\o_a)(\half\D^af\pm\o^a)+\D^a(\half\D_af\pm\o_a)\right)
=8\pi\rho&(3b)\cr}$$
where $\phi_\pm=T(u_\pm-r_\pm,u_\pm-r_\pm)$
and $\rho=T(u_+-r_+,u_--r_-)$
in terms of the energy tensor $T$,
with units $G=1$.
These are the focussing equations,
with (3a) describing how neighbouring light rays focus in each null congruence,
and (3b) describing the cross-focussing between the two null congruences.
\ms
The dominant energy condition [3] implies
$$\eqalignno
{&\phi_\pm\ge0&(4a)\cr
&\rho\ge0.&(4b)\cr}$$
Condition (4a) requires only the null convergence condition [3],
referred to subsequently as the null energy condition,
which follows from the weak energy condition [3].
All quantities will be assumed to exist as functions on $S$,
and in particular an energy tensor with distributional support is disallowed,
since this can produce artificial discontinuous jumps in the trapping horizon.
Rather remarkably, the focussing equations (3) and energy inequalities (4)
suffice to determine the key properties of trapping horizons.
\bs\ni
{\bf V. Topology, signature and area}
\ms\ni
There are two relevant integral theorems,
discussed in this context by Newman [10].
Firstly, if a vector $X$ tangent to $S$ is such that
$\D_aX^a$ and $|X|=\sqrt{X_aX^a}$ are integrable on $S$, then
$$\I\D_aX^a=0.\eqno(5)$$
For compact $S$, this is just the Gauss theorem [7].
Secondly, if $\R$ is integrable on $S$,
there is the Cohn-Vossen inequality [10],
$$\I\R\le4\pi\chi\eqno(6)$$
where $\chi$ is the Euler-Poincar\'e characteristic of $S$,
with $\chi=2$ for a sphere, $\chi=1$ for a plane or projective plane,
$\chi=0$ for a torus, cylinder, Klein surface or M\"obius band,
and $\chi\le-1$ for any other surface.
For compact orientable 2-surfaces,
the Cohn-Vossen inequality reduces to an equality,
the Gauss-Bonnet theorem [7].
If $\R$,
$\D^a(\o_a-\half\D_af)$ and $|\o-\half\D f|$
are integrable on an outer or inner marginal surface $S$,
then $S$ will be described as {\it well adjusted}.
This condition can be thought of as expressing that
a non-compact $S$ is asymptotically well behaved,
cf.\ `well tempered' [10].
A compact marginal surface is necessarily well adjusted.
\ms\ni
{\it Theorem 1: topology law.}
If the dominant energy condition holds,
a well adjusted, future or past, outer marginal surface
has spherical or planar topology.
\ms\ni
{\it Proof.}
For an outer marginal surface $S$,
the cross-focussing equation (3b) yields the inequality
$0<\O{-\hbox{e}^f\L_-\t_+}\le\O{\half\R}+\O{\D^a(\half\D_af-\o_a)}$,
which integrates over well adjusted $S$ to $0<2\pi\chi$,
using (5) and (6).
Since future or past marginal surfaces must be orientable,
this leaves only spherical ($\chi=2$) or planar ($\chi=1$) topology.
\ms\ni
{\it Corollary.}
If the dominant energy condition holds,
a compact, future or past, outer marginal surface has spherical topology.
\ms\ni
Note that there is no such restriction
on the topology of inner marginal surfaces.
The result generalizes Hawking's `sphere theorem'
for asymptotically flat space-times [2--3].
Hawking considered the outermost apparent horizon,
i.e.\ the apparent horizon nearest to conformal infinity,
which must be an outer rather than inner trapping horizon,
due to the signs of the expansions near infinity.
The result is similar to that of Newman~[10],
whose `stable marginal surfaces' are spheres, planes or projective planes.
Note that if degenerate outer marginal surfaces are allowed,
$\O{\L_-\t_+}\le0$,
then $\chi=0$ topologies are also possible.
However, (6) and (3b) then show that
$\O\rho$, $\O{(\o-\half\D f)}$ and $\O{\L_-\t_+}$ vanish,
and hence so does $\O\R$.
Such flat surfaces can be dismissed on the grounds of genericity or stability.
\ms\ni
{\it Theorem 2: signature law.}
If the null energy condition holds,
a trapping horizon is null if and only if
the internal shear and normal energy density vanish.
Otherwise, an outer trapping horizon is spatial
and an inner trapping horizon is Lorentzian.
\ms\ni
{\it Proof.}
The direction of development of the trapping horizon $\C$
is located by a vector $\H z$
tangent to $H$ and normal to the foliating marginal surfaces,
such that $\H{\L_z\t_+}=0$.
So $z$ is a linear combination $z=\H{\b(u_+-r_+)}-\H{\a(u_--r_-)}$,
and is spatial, null or temporal
as $\a/\b$ is positive, zero or negative respectively. Since
$\H{\L_z\t_+}=\H{\b\L_+\t_+}-\H{\a\L_-\t_+}$,
one has $\a/\b=\H{\L_+\t_+/\L_-\t_+}$.
The focussing equation (3a) yields $\H{\L_+\t_+}\le0$,
so that $\a/\b\ge0$ for an outer trapping horizon
and $\a/\b\le0$ for an inner trapping horizon.
In both cases, $\a/\b|_p=0$ at $p\in H$ if and only if $\L_+\t_+|_p=0$,
i.e.\ in the `instantaneously stationary' case
where the normal energy density $\phi_+|_p$
and the internal shear $\s_+|_p$ vanish.
The result follows from the sign of $\a/\b$.
\ms\ni
This verifies the common belief that an outer trapping horizon,
or something similar like an apparent horizon, is generically spatial.
Again, this is not true for inner trapping horizons.
Newman [10] gave a definition of stability for marginal surfaces
in terms of the existence of a variation which is spatial or null.
Newman's `stable marginal surface' is thus closely related
to the outer marginal surface as defined here.
In this sense, inner trapping horizons are unstable.
(Except, ironically, for instantaneously stationary inner trapping horizons).
Although this definition of stability is questionable,
it relates to the general belief that inner black-hole horizons are unstable
in the usual initial-data sense [4].
This raises the question of whether
there is some reason of principle which forbids inner horizons.
\ms\ni
{\it Theorem 3: second law.}
If the null energy condition holds,
the area form of a future outer or past inner trapping horizon
is non-decreasing,
and the area form of a past outer or future inner trapping horizon
is non-increasing,
in all cases being constant if and only if the horizon is null.
\ms\ni
{\it Proof.}
Note that $\L_\pm\mu=\mu\t_\pm$.
Along the horizon,
the area form $\mu$ develops according to
$\H{\L_z\mu}=\H{-\mu\a\t_-}$.
The orientation of $z$ is naturally fixed by $\b>0$,
meaning that $z$ has positive component
along the future-pointing null direction of vanishing expansion.
Then outer trapping horizons have $\a\ge0$
and inner trapping horizons $\a\le0$,
and by definition future trapping horizons have $\H{\t_-}<0$
and past trapping horizons $\H{\t_-}>0$.
\ms\ni
{\it Corollary.}
The area form of a future trapping horizon is increasing, constant or
decreasing
as the horizon is spatial, null or Lorentzian respectively.
\ms\ni
The second law, or area theorem, is the analogue for trapping horizons,
conjectured by Hawking [2] for apparent horizons
in the asymptotically flat context,
of his famous theorem for event horizons.
Note that the second law is a pointwise result,
with the immediate consequence that, for compact marginal surfaces $S$,
the total area
$$A=\I\eqno(7)$$ has the same monotonicity properties.
For outer trapping horizons, this confirms the expectation that
black holes have non-decreasing area,
and white holes have non-increasing area.
For inner trapping horizons, the situation reverses.
Considering the thermodynamic analogy,
this reversal of the thermodynamic arrow of time on an inner horizon
could be taken as indicating either novel physics,
or that black holes with inner horizons cannot occur in nature
for thermodynamic reasons.
\bs\ni
{\bf VI. Trapping gravity and the zeroth and first laws}
\ms\ni
As already noted, the area theorem for a future outer trapping horizon
is analogous to the second law of thermodynamics,
with area $A$ corresponding to entropy $s\sim{1\over4}A$.
To formulate analogues of the other laws of thermodynamics
requires an analogue of temperature $\vartheta$,
namely surface gravity $\k\sim2\pi\vartheta$.
Surface gravity has previously been defined
only for Killing horizons, i.e.\ for stationary black holes,
and the standard zeroth and first laws are formulated for this case only [1].
In seeking a general definition of surface gravity,
one must choose between the event horizon and the trapping horizon,
which coincide only in the stationary case.
Since the event horizon is an essentially global concept,
there can be no local or quasi-local definition of surface gravity
associated with it.
For a trapping horizon, a simple geometrical definition is suggested by
recalling that a basic property of surface gravity is
that its vanishing is the condition for the horizon to be degenerate.
\ms\ni
{\it Definition.}
The {\it trapping gravity} $\k$ of an outer trapping horizon $\C$
is given by
$$\k=\H{\half\sqrt{-\hbox{e}^f\L_-\t_+}}
=\H{\half\sqrt{-\hbox{e}^f\mu^{-1}\L_-\L_+\mu}}
\eqno(8)$$
where the double-null foliation is adapted to the marginal surfaces.
\ms\ni
Recall from Section II that $\k$ is an invariant of a non-null trapping
horizon.
Note that $\k>0$,
which may be relaxed to $\k\ge0$ if one wishes to include degenerate horizons.
The supposed `third law of black-hole dynamics', namely $\k$ real and positive,
is then linked to the supposed non-existence of inner or degenerate horizons.
It transpires that there is a connection between $\k$
and various quasi-local measures of energy.
\ms\ni
{\it Definitions.}
On a trapping horizon foliated by compact marginal surfaces $S$,
the irreducible energy $m$ [11],
angular energy $a$ and material energy $q$
are given by
$$\eqalignno{
&m=\sqrt{{A\over{16\pi}}}&(9a)\cr
&a=m\sqrt{{1\over{4\pi}}\I(\o_a-\half\D_af)(\o^a-\half\D^af)}&(9b)\cr
&q=m\sqrt{2\I\hbox{e}^f\rho}.&(9c)\cr}$$
\ni
These are invariants of the trapping horizon,
with $m$ being a convenient measure of the area,
$a$ depending on the extrinsic curvature,
and $q$ depending on the energy tensor of the matter.
The energy or mass $m$ is irreducible in the sense that
it is non-decreasing on a future outer trapping horizon,
according to the second law.
The quantity $ma$ may be interpreted as measuring quasi-local angular momentum,
and is related to the Hamiltonian quasi-local energy [12],
which, if $\O{(f,r_\pm,\nu_\pm)}=0$, takes the form
$$E={m\over{8\pi}}\I\left(\R+\t_+\t_--\half\s^+_{ab}\s_-^{ab}-2\o_a\o^a\right).
\eqno(10)$$
Note that $E$ includes a contribution from the angular term.
\ms\ni
{\it Theorem 4: zeroth law.}
If the dominant energy condition holds
on a compact, future or past, outer marginal surface $S$,
the total trapping gravity is bounded above according to
$$\I\k\le4\pi\sqrt{m^2-a^2-q^2}\eqno(11)$$
with equality if and only if $\k$ is constant on $S$.
\ms\ni
{\it Proof.}
On an outer marginal surface $S$,
the cross-focussing equation (3b) reduces to
$$4\k^2=\O{\left(\half\R-(\half\D_af-\o_a)(\half\D^af-\o^a)
+\D^a(\half\D_af-\o_a)-8\pi\hbox{e}^f\rho\right)}\eqno(12)$$
which integrates to
$$\I\k^2=\pi\left(1-{a^2\over{m^2}}-{q^2\over{m^2}}\right)\eqno(13)$$
recalling from the topology law that $S$ must have spherical topology,
and that $\I\R=8\pi$.
The result follows from the Cauchy-Schwarz inequality,
$(\I\varphi)^2\le A\I\varphi^2$,
with equality if and only if $\varphi$ is constant on $S$.
\ms\ni
This zeroth law is a much stronger statement
than the standard version for stationary horizons,
which merely states that the surface gravity is constant in that case.
The new version gives a necessary and sufficient condition
for the trapping gravity to be constant,
and shows that the equilibrium case corresponds to the attainment of a bound.
This provides a powerful theoretical reason
for regarding the equilibrium case as physically important,
rather than just a convenient simplification.
An analogous statement in thermodynamics would be that
there is an upper bound on the total temperature of a system,
attained if and only if the temperature is constant,
i.e\ in thermal equilibrium.
\ms
In the equilibrium case, the inequality reduces to
$A\k=4\pi\sqrt{m^2-a^2-q^2}$,
which has the same form as
the usual formula for the surface gravity of the Kerr-Newman solution,
though be warned that $(m,a,q)$ are quasi-local quantities
defined on the horizon rather than the traditional asymptotic quantities.
Note that this is a dynamic equilibrium
in the sense that $\k$ need not be constant along the horizon,
i.e.\ the black hole need not be stationary.
More specifically still, the inequality (11) also implies the maximal bound
$\I\k\le4\pi m$,
which is attained only if
$\O\rho$ and $\O{(\o-\half\D f)}$ vanish and $\O\R$ is constant,
so that $S$ is a metric sphere with $4m\k=1$.
This is still more general than
the case of the Schwarzschild horizon,
since infalling matter can be present, i.e.\ the horizon need not be null.
An example is the massless scalar field,
for which $\rho$, $\o$ and $\D f$ vanish in spherical symmetry.
%Variations among this class satisfy $\k\,\d A=8\pi\,\d m$.
\ms
Given the generalization of surface gravity to trapping gravity,
consider generalizing the first law of black-hole dynamics,
which should relate variations in the area of the horizon
to variations in energy, analogously to $\vartheta\,\d s=\d\e$.
In the context of Killing horizons,
the variation $\d$ is usually taken among Kerr-Newman solutions,
parametrized by the asymptotic constants.
For a trapping horizon,
the natural variation to consider is just the derivative along the horizon.
\ms\ni
{\it Theorem 5: first law.}
If the null energy condition holds
on an outer trapping horizon $\C$,
the area form $\mu$ of the marginal surfaces develops according to
$$\k\H{\L_z\mu}=8\pi\Phi
=\H{-\mu\hbox{e}^f\t_-
\sqrt{\pi\phi_++\tt\s^+_{ab}\s_+^{ab}}}\eqno(14)$$
where the vector $z$ is tangent to the horizon and normal to the surfaces,
and normalized by $z_az^a=1$ if spatial.
\ms\ni
{\it Proof.}
Recall from the signature law that on an outer trapping horizon,
$z$ is spatial or null,
being null at $p\in H$ if and only if $\L_z\mu|_p=0$,
in which case the result is trivial.
When $z$ is spatial, it may be normalized to a unit vector by
$z=\H{((u_+-r_+)\hbox{e}^f/2\a-\a(u_--r_-))}$.
Then $0=\H{\L_z\t_+}=\H{(\L_+\t_+\hbox{e}^f/2\a-\a\L_-\t_+)}$
yields $2\a^2=\H{\hbox{e}^f\L_+\t_+/\L_-\t_+}$,
with $\H{\L_+\t_+}=\H{-(8\pi\phi_++\quart\s^+_{ab}\s_+^{ab})}$
and $\H{\L_-\t_+}=-4\hbox{e}^{-f}\k^2$.
The result follows from $\H{\L_z\mu}=\H{-\mu\a\t_-}$.
\ms\ni
This new version of the first law applies to any outer trapping horizon,
and refers to the actual variation of the area form along the horizon,
rather than a variation in some space of solutions.
Remarkably, the trapping gravity occurs in the expected way.
Note that this is another pointwise result,
with the variation of the total area being obtained as the integral,
$$\L_zA=\int_S{8\pi\Phi\over\k}.\eqno(15)$$
The 2-form $\Phi$ is not immediately familiar,
but may be interpreted as measuring the effective energy flux
across the horizon,
with the factor of $\hbox{e}^f\t_-$ ensuring that
$\Phi$ is invariant under rescalings of the null normals.
Alternatively, one may avoid any such interpretation
and simply regard the result as determining
the development of the area form in terms of initial data on the horizon.
The main point is that if there is a general first law for trapping horizons,
equation (14) is the correct expression.
\ms
The new first law can be written as a balance of variations
by defining a mass aspect or energy 2-form $\e$ such that $\Phi=\L_z\e$.
The constant of integration could be fixed
by taking $\e$ zero at the centre where the trapping horizon first forms,
presuming this exists.
This in turn suggests defining the energy or mass of a trapping horizon
up to a marginal surface $S$ by $\E=\int_S\e$.
Note that $\E$ is implicitly an integral over the horizon,
rather than an integral over $S$ alone.
\bs\ni
{\bf VII. Conclusion}
\ms\ni
Several key properties of trapping horizons have been proved,
including general versions of
the zeroth, first and second laws of black-hole dynamics.
At a more fundamental level, the {\it future outer trapping horizon}
provides a general definition of a black hole.
The zeroth and first laws involve
a general definition of the surface gravity of a black hole,
the trapping gravity.
A genuine connection to thermodynamics, namely Hawking radiation,
is known only for the stationary case [13--14].
How or whether the phenomenon of Hawking radiation generalizes
to trapping horizons is currently an important open question.
\bs\ni
Acknowledgements.
It is a pleasure to thank Dieter Brill, Marcus Kriele, Ken-ichi Nakao,
Richard Newman, Tetsuya Shiromizu, Paul Tod and James Vickers
for discussions that led to some improvements in the preprint.
I am also grateful to
the Max-Planck-Gesellschaft for initial financial support,
and the Southampton relativity group for subsequent hospitality.
\np\ni
{\bf Appendix A: relating marginal surfaces and trapping boundaries}
\ms\ni
The main motivation for considering marginal surfaces is that
since they are `just trapped',
they should give a quasi-local characterization of
the boundary of a black or white hole.
This appendix shows that compact,
future or past, outer or inner marginal surfaces
lie in closures of trapped regions,
and are the only non-degenerate marginal surfaces which
foliate boundaries of inextendible trapped regions.
\ms\ni
{\it Definitions.}
A {\it trapped region} of space-time $\M$ is a connected subset $T\subseteq\M$
such that for all $p\in T$,
there exists a trapped surface $S'$ with $p\in S'$.
An {\it extension} $T'$ of $T$ is a trapped region with $T\subset T'$.
An {\it inextendible} trapped region is one with no extension.
A {\it trapping boundary} is a connected component of the boundary
of an inextendible trapped region.
\ms\ni
This is inspired by Hawking's definition of apparent horizon [2--3]
but differs in several respects.
Firstly,
trapped surfaces rather than `outer' trapped surfaces have been used,
since the inside and outside of a 2-surface
cannot be distinguished in general.
(The classical work refers to asymptotically flat space-times,
where such a distinction can be made).
Secondly, the trapped regions are 4-dimensional rather than 3-dimensional,
for maximal generality.
Thirdly, there are no additional global assumptions,
such as `regular predictability'.
\ms\ni
{\it Theorem 6.}
Given a compact marginal surface $S$ lying in
a future or past, outer or inner trapping horizon,
any spatial 2-surface sufficiently close to $S$
is trapped if it lies inside the horizon,
and not trapped if it lies outside.
\ms\ni
{\it Proof.}
It suffices to consider one of the four cases,
say a future outer trapping horizon,
so that on one of the foliating marginal surfaces $S$,
$\O{\t_+}=0$, $\O{\t_-}<0$ and $\O{\L_-\t_+}<0$.
Note that $\t_-<0$ in a neighbourhood of $S$.
Consider any 3-surface $\Sigma$ which crosses the horizon at $S$.
The direction $\O v$ tangent to $\Sigma$ and normal to $S$
can be written as a linear combination $v=\O{\g(u_--r_-)}+\O{\d z}$,
where $\O z=\O{\b(u_+-r_+)}-\O{\a(u_--r_-)}$ is the direction of the horizon,
given by $\O{\L_z\t_+}=0$,
and $\g>0$ fixes the orientation of $v$.
Then $\O{\L_v\t_+}=\O{\g\L_-\t_+}<0$,
so that for any foliation of $\Sigma$ based on $S$,
there are neighbourhoods $U$ and $U'$ of $\Sigma$
with $\partial U\cap\partial U'=S$,
inside and outside the horizon respectively,
such that $\t_+|_U<0$
and $\t_+|_{U'}>0$,
and $\t_+\t_-|_U>0$
and $\t_+\t_-|_{U'}<0$.
Since $\Sigma$ and the foliation of $\Sigma$ are arbitrary,
any spatial 2-surface sufficiently close to $S$
is trapped if it lies to the future of the horizon,
and not trapped if it lies to the past.
\ms\ni
{\it Corollary.}
A compact, future or past, outer or inner marginal surface
lies in the closure of a trapped region.
\ms\ni
It does not necessarily follow that
the marginal surface lies in a trapping boundary,
since it could still be intersected by a non-neighbouring trapped surface.
For instance, when black holes collide,
the trapped regions merge and the trapping boundary changes topology.
However, marginal surfaces are likely to continue into the trapped region,
forming a ghostly trapping horizon.
\ms
Establishing a converse is more technical.
In principle, a trapping boundary may not be smooth,
and the trapped surfaces may be distributed chaotically near the boundary.
It seems that further progress requires assumptions
on the smoothness of the boundary
and on the uniformity with which trapped surfaces approach the boundary.
\ms\ni
{\it Definition.}
A {\it limit section} of a trapping boundary is a subset
which is a spatial 2-surface
arising as the uniform limit of a sequence of trapped surfaces.
\ms\ni
Assuming that a trapping boundary admits such limit sections
is probably a physically reasonable requirement,
but from a purely mathematical viewpoint is a rather strong condition.
In fact,
considerably weaker conditions suffice for the laws derived in this article,
or pointwise versions thereof, but the proofs become significantly longer.
(I particularly thank Marcus Kriele for discussions on these issues).
\ms\ni
{\it Theorem 7.}
If a subset of a trapping boundary can be foliated by limit sections $S$, then
\item{(i)} $\O{\t_+\t_-}=0$, and taking $\O{\t_+}=0$,
\item{(ii)} either $\O{\t_-}\le0$ or $\O{\t_-}\ge0$, and
\item{(iii)} either $\O{\L_-\t_+}\le0$ or $\O{\L_-\t_+}\ge0$.
\ms\ni
{\it Proof of (i).}
Taking a sequence of trapped surfaces $S_i$ approaching $S$,
$\t_+\t_-|_{S_i}>0$ implies $\O{\t_+\t_-}\ge0$.
Next suppose that there is a $q\in S$ such that $\t_+\t_-|_q>0$.
Then for any foliation there exists a neighbourhood $U$ of $q$
such that $\t_+\t_-|_U>0$.
Consider a foliation containing a trapped surface $S'$
passing through $U$
and a surface $S''$ which crosses $S$ in $U$
and is arbitrarily close to $S'$ outside $U$.
Then $\t_+\t_-|_{S''}>0$, so that $S''$ is also trapped.
This contradicts the inextendibility of the trapped region.
Thus $\O{\t_+\t_-}=0$.
\ms\ni
{\it Proof of (ii).}
On the sequence $S_i$, either $\t_-|_{S_i}<0$ or $\t_-|_{S_i}>0$,
corresponding to future and past trapped surfaces respectively.
So either $\O{\t_-}\le0$ or $\O{\t_-}\ge0$ respectively.
\ms\ni
{\it Proof of (iii).}
On the sequence $S_i$, either $\t_+|_{S_i}<0$ or $\t_+|_{S_i}>0$,
corresponding to future and past trapped surfaces respectively.
Since $\O{\t_+}=0$,
either $\O{\L_v\t_+}\le0$ or $\O{\L_v\t_+}\ge0$ respectively,
where $\O v$ is the direction normal to $S$
along which the sequence approaches $S$, pointing towards the trapped region.
The direction $\O z$ of the trapping boundary, for which $\O{\L_z\t_+}=0$,
can then be related to $v$ by $v=\O{\g(u_--r_-)}+\O{\d z}$,
with $\g>0$ or $\g<0$
as $S_i$ are to the future or past of the horizon respectively.
Since $\O{\L_v\t_+}=\O{\g\L_-\t_+}$,
either $\O{\L_-\t_+}\le0$ or $\O{\L_-\t_+}\ge0$.
\ms\ni
The main lesson of (ii) and (iii) is that
neither $\t_-$ nor $\L_-\t_+$ can vary its sign over $S$,
thus classifying marginal surfaces that occur in trapping boundaries
into four basic types,
described as future or past, and outer or inner,
in view of the example of the Kerr or Reissner-Nordstr\"om black hole.
The non-strict inequalities in (ii) and (iii)
have been made strict in the definition given in the text,
because the converses of (ii) and (iii) are not true,
i.e.\ a marginal surface satisfying (ii) and (iii)
need not lie in a trapping boundary,
as shown by the example of marginal spheres in the de Sitter cosmos,
or marginal planes in plane-wave space-times.
So the definition of `future outer marginal surface'
excludes some degenerate marginal surfaces
which may also lie in a trapping boundary.
The degenerate cases have been excluded
partly because they are more awkward for calculational purposes,
and partly due to the putative physical principle,
referred to as the `third law of black-hole dynamics',
that degenerate horizons cannot be attained by physical processes.
\np\ni
{\bf Appendix B: anholonomicity vs.\ normal fundamental form}
\ms\ni
In classical differential geometry,
the extrinsic curvature of a submanifold is encoded in
the second fundamental forms and the normal fundamental forms [7].
It is important to note that the normal fundamental forms
are not genuine geometrical invariants,
but depend on the choice of basis of the normal bundle.
Fortunately,
there is a geometrical invariant, the anholonomicity or `twist' [5--6],
which encodes essentially the same information.
In general,
the anholonomicity is an invariant associated with a particular foliation,
but in the case of co-signature $(-+)$,
the null normals fix a preferred normal basis,
and the value of the anholonomicity on a particular submanifold
turns out to be independent of the foliation away from the submanifold.
\ms
Consider a manifold with metric $\con\cdot\cdot$,
and a submanifold $S$ of codimension $m$.
Take a normal basis $N_\mu=(N_1,\ldots N_m)$,
i.e.\ a linearly independent set of vectors such that
$\con{N_\mu}{X}=0$ for all vectors $X$ tangent to $S$.
The {\it normal fundamental forms} $\b_{\mu\nu}$
are 1-forms on $S$ defined by
$$\b_{\mu\nu}(X)=\con{\N_XN_\mu}{N_\nu}.$$
Clearly $\b_{\mu\nu}$ are not independent of the choice of basis $N_\mu$.
If the co-signature is Riemannian,
it is usual to take $N_\mu$ to be orthonormal,
but in general it is not possible to maintain such a normalization
over a foliation of submanifolds.
\ms
If the co-signature is $(-+)$,
there are two preferred normal directions, the null directions,
given by vectors $N_+$ and $N_-$ such that $\con{N_\pm}{N_\pm}=0$.
This fixes a preferred basis for the normal bundle,
up to rescalings and interchange of $N_\pm$.
There are then only two non-zero normal fundamental forms,
$$\b_\pm(X)=\con{\N_XN_\pm}{N_\mp}.$$
Under rescalings $N_\pm\mapsto\hbox{e}^{c_\pm}N_\pm$,
$$\b_\pm\mapsto
\hbox{e}^{c_+}\hbox{e}^{c_-}\left(\b_\pm+\con{N_+}{N_-}\N_{(S)}c_\pm\right)$$
and under interchange $N_\pm\mapsto N_\mp$,
$$\b_\pm\mapsto\b_\mp.$$
Interpreting, $\b_\pm$ are not invariants of the extrinsic curvature,
but depend on the choice of normals.
This can be corrected in three steps.
(i) Normalize $\b_\pm$ by dividing by $\con{N_+}{N_-}$.
(ii) Antisymmetrize, i.e.\ use $\b_+-\b_-$ as the variable,
since $\b_++\b_-=\N_{(S)}\con{N_+}{N_-}$ encodes only coordinate data.
(iii) Propagate $N_\pm$ such that
the corresponding 1-forms $n_\pm(z)=\con{N_\pm}{z}$ are closed,
$\hbox{d}n_\pm=0$.
The resulting 1-form on $S$ is the {\it anholonomicity} $\o$,
$$\o(X)={\con{[N_-,N_+]}{X}\over{2\con{N_+}{N_-}}}
={\con{\N_XN_+}{N_-}-\con{\N_XN_-}{N_+}\over{2\con{N_+}{N_-}}}.$$
This time, under rescalings $N_\pm\mapsto\hbox{e}^{c_\pm}N_\pm$
preserving $\hbox{d}n_\pm=0$,
$$\o\mapsto\o$$
and under interchange $N_\pm\mapsto N_\mp$,
$$\o\mapsto-\o.$$
So $|\o|$ is an invariant of the extrinsic curvature.
The condition $\hbox{d}n_\pm=0$
relates the normal fundamental forms to the anholonomicity by
$\b_\pm=\half\N_{(S)}\con{N_+}{N_-}\pm\con{N_+}{N_-}\o$.
On a particular submanifold $S$,
one may fix $\O{\con{N_+}{N_-}}=-1$,
in which case $\O{\b_\pm}=\O{\mp\o}$,
so that the normal fundamental forms coincide with the anholonomicity on $S$,
up to sign.
This means that the distinction is irrelevant for many purposes.
However, $\con{N_+}{N_-}$ cannot be freely prescribed over a foliation,
and so the invariant $\o$
is generally preferable to the basis-dependent normal fundamental forms.
\ms
More generally, the anholonomicity is defined
with respect to a foliation of submanifolds as follows.
Consider $m$ foliations of codimension-1 submanifolds
given by the level surfaces of $m$ functions $\xi_\mu=(\xi_1,\ldots\xi_m)$,
which intersect in a foliation of codimension-$m$ submanifolds.
Define the normal 1-forms $n_\mu=\hbox{d}\xi_\mu$
and the corresponding normal vectors $N_\mu$,
$\con{N_\mu}{z}=n_\mu(z)$.
The anholonomicity $\o_{\mu\nu}$ is defined by
$$\o_{\mu\nu}(X)=\half\con{[N_\nu,N_\mu]}{X}
=\half\con{\N_XN_\mu}{N_\nu}-\half\con{\N_XN_\nu}{N_\mu}$$
up to normalization, the details of which depend on the co-signature.
Comparing with the normal fundamental forms:
on at least one (codimension-$m$) submanifold $S$,
the symmetric part $\b_{(\mu\nu)}=\half\N_{(S)}\con{N_\mu}{N_\nu}$
can be set to zero by coordinate choice,
and the antisymmetric part $\b_{[\mu\nu]}$ coincides with $\o_{\mu\nu}$
as a result of $\hbox{d}n_\mu=0$.
A suitably normalized $\o_{\mu\nu}$
is invariant under functional rescalings of $\xi_\mu$,
so is an invariant of the foliation.
Its value on $S$ is independent of the foliation away from $S$
only in the case of co-signature $(-+)$,
where the null normals fix preferred normal directions.
Otherwise, rotations of the normal basis preclude
the extraction of preferred 1-forms from $\o_{\mu\nu}$.
\np
\begingroup
\parindent=0pt\everypar={\global\hangindent=20pt\hangafter=1}\par
{\bf References}\ms
\refb{[1] Carter B 1973 in}{Black Holes}
{ed: DeWitt C \& DeWitt B S (Gordon \& Breach)}
\refb{[2] Hawking S W 1973 in}{Black Holes}
{ed: DeWitt C \& DeWitt B S (Gordon \& Breach)}
\refb{[3] Hawking S W \& Ellis G F R 1973}
{The Large Scale Structure of Space-Time}{(Cambridge University Press)}
\refb{[4] Penrose R 1968 in}{Battelle Rencontres}
{ed: DeWitt C M \& Wheeler J A (Benjamin)}
\ref{[5] d'Inverno R A \& Smallwood J 1980}\PR{D22}{1233}
\ref{[6] Hayward S A 1993}\CQG{10}{779}
\refb{[7] Spivak M 1979}{A Comprehensive Introduction to Differential Geometry}
{(Publish or Perish)}
\ref{[8] Tipler F J 1977}{Nature}{270}{500}
\refb{[9] Penrose R \& Rindler W 1986 \& 1988}
{Spinors and Space-Time Volumes 1 \& 2}{(Cambridge University Press)}
\ref{[10] Newman R P A C 1987}\CQG4{277}
\ref{[11] Christodoulou D \& Ruffini R 1971}\PR{D4}{3552}
\ref{[12] Hayward S A 1994}\PR{D49}{831}
\ref{[13] Hartle J B \& Hawking S W 1976}\PR{D13}{2188}
\ref{[14] Gibbons G W \& Hawking S W 1977}\PR{D15}{2738}
\endgroup
\bye